\documentclass[fleqn,usenatbib]{mnras}

\usepackage{newtxtext,newtxmath}

\usepackage[T1]{fontenc}

\DeclareRobustCommand{\VAN}[3]{#2}
\let\VANthebibliography\thebibliography
\def\thebibliography{\DeclareRobustCommand{\VAN}[3]{##3}\VANthebibliography}

\usepackage{graphicx}	
\usepackage{amsmath}	

\title[Lensed star transient in the Cosmic Horseshoe]{Lensed stars in galaxy-galaxy strong lensing -- a JWST prediction for the Cosmic Horseshoe}

\author[S.K.Li et al.]{
Sung Kei Li,$^{1}$\thanks{E-mail: keihk98@connect.hku.hk} Luke Weisenbach,$^{2}$ Thomas E. Collett,$^{2}$ Jose M. Diego,$^{3}$ Jeremy Lim,$^{1}$
\newauthor{Thomas J. Broadhurst,$^{4}$$^{,5}$$^{,6}$  Alex Chow,$^{1}$ Wolfgang J.R. Enzi,$^{2}$  Patrick L. Kelly,$^{7}$  Carlos R. Melo-Carneiro,$^{8}$
  }
\newauthor{Jose M. Palencia,$^{3}$ Liliya L.R. Williams,$^{7}$ Jiashuo Zhang$^{1}$}
\\
$^{1}$Department of Physics, The University of Hong Kong, Pokfulam Road, Hong Kong\\
$^{2}$Institute of Cosmology and Gravitation, University of Portsmouth, Burnaby Road, Portsmouth PO1 3FX, UK\\
$^{3}$Instituto de Física de Cantabria (CSIC-UC), Avda. Los Castros s/n, 39005 Santander, Spain \\
$^{4}$Department of Theoretical Physics, University of Basque Country UPV/EHU, Bilbao, Spain \\
$^{5}$Ikerbasque, Basque Foundation for Science, Bilbao, Spain \\
$^{6}$Donostia International Physics Center, Paseo Manuel de Lardizabal, 4, San Sebasti\'an, 20018, Spain\\
$^{7}$Minnesota Institute for Astrophysics, University of Minnesota, 116 Church St. SE, Minneapolis, MN 55455, USA\\
$^{8}$Instituto de Física, Universidade Federal do Rio Grande do Sul, Av. Bento Gonçalves 9500, Porto Alegre-RS, 90040-060, Brazil 
}

\date{Accepted 2025 November 5. Received 2025 November 5; in original form 2025 September 20}

\pubyear{\the\year{}}

\begin{document}
\label{firstpage}
\pagerange{\pageref{firstpage}--\pageref{lastpage}}
\maketitle

\begin{abstract}

We explore for the first time the possibility of detecting lensed star transients in galaxy-galaxy strong lensing systems upon repeated, deep imaging using the {\it James-Webb Space Telescope} ({\it JWST}). Our calculation predicts that the extremely high recent star formation rate of $\sim 140\,M_{\odot}\textrm{yr}^{-1}$ over the last 50 Myr (not accounting for image multiplicity) in the ``Cosmic Horseshoe'' lensed system ($z = 2.381$) generates many young, bright stars, of which their large abundance is expected to lead to a detection rate of $\sim 60$ transients per pointing in {\it JWST} observations with a $5\sigma$ limiting magnitude of $\sim 29\,m_{AB}$. With the high expected detection rate and little room for uncertainty for the lens model compared with cluster lenses, our result suggests that the Cosmic Horseshoe could be an excellent tool to test the nature of dark matter based on the spatial distribution of transients, and can be used to constrain axion mass if dark matter is constituted of ultra-light axions.
We also argue that the large distance modulus of $\sim46.5\,$mag at $z \approx 2.4$ can act as a filter to screen out less massive stars as transients and allow one to better constrain the high-mass end of the stellar initial mass function based on the transient detection rate. Follow-up {\it JWST} observations of the Cosmic Horseshoe would allow one to better probe the nature of dark matter and the star formation properties, such as the initial mass function at the cosmic noon, via lensed star transients.

\end{abstract}





\section{Introduction}


Lensed stars at redshifts $\gtrsim 1$ have been regularly detected in galaxies that are strongly lensed by galaxy clusters since their first discovery \citep{Kelly_2018_Icarus, Rodney_2018, Chen_2019_Warhol, Kelly_2022_Flashlights, Meena_2023_Flashlights, Meena_2023_MACSJ0647, Yan_2023, Fudamoto_2025}. First predicted by \citet{Miralda-Escude_1991}, it is the combination of strong-lensing (from galaxy clusters) and microlensing (from stars such as intracluster stars) up to a magnification factor of $\sim 10^{4}$ that allows individual, luminous stars at cosmological distances to be detected temporarily \citep{Oguri_2018, Diego_2019_extrememagnification, 2024SSRv..220...57W}. The increasing amount of detection has introduced a new era of astronomy, allowing the study of various astrophysical topics with the statistics of these events. This includes studying the nature of dark matter based on the spatial distribution of these transient events \citep{Williams_2024, Broadhurst_2025}, the abundance of massive stars and thus the recent star formation history of lensed galaxies \citep{Diego_2024_buffaloflashlights, Li_2025_BRratio}, and even constraining the stellar initial mass function (IMF) at the earlier universe \citep{Li_2025_IMF, Meena_2025_IMF, Williams_2025}.

To date, all lensed stars have been discovered towards galaxy clusters, with no single event associated with galaxy-galaxy strong lensing (GGSL) systems. 
Given that the mass distribution in GGSLs can often be better modelled (in terms of the capability of reproducing the image of lensed galaxies) than the case in galaxy clusters \citep[see recent review of the divergence of cluster modelling in][]{Perera_2025} despite many recent efforts of source reconstruction in cluster lenses \citep[e.g.,][]{Acebron_2024, Xie_2025, galan2025stronglensingmodeldust}, studying transients in GGSL systems mitigates uncertainties arising from the adopted lens models \citep{Broadhurst_2025, Li_2025_IMF}. This further motivates us to investigate whether it is possible to detect transients in GGSLs in this paper.

\begin{figure}
    \centering
    \includegraphics[width=1\linewidth]{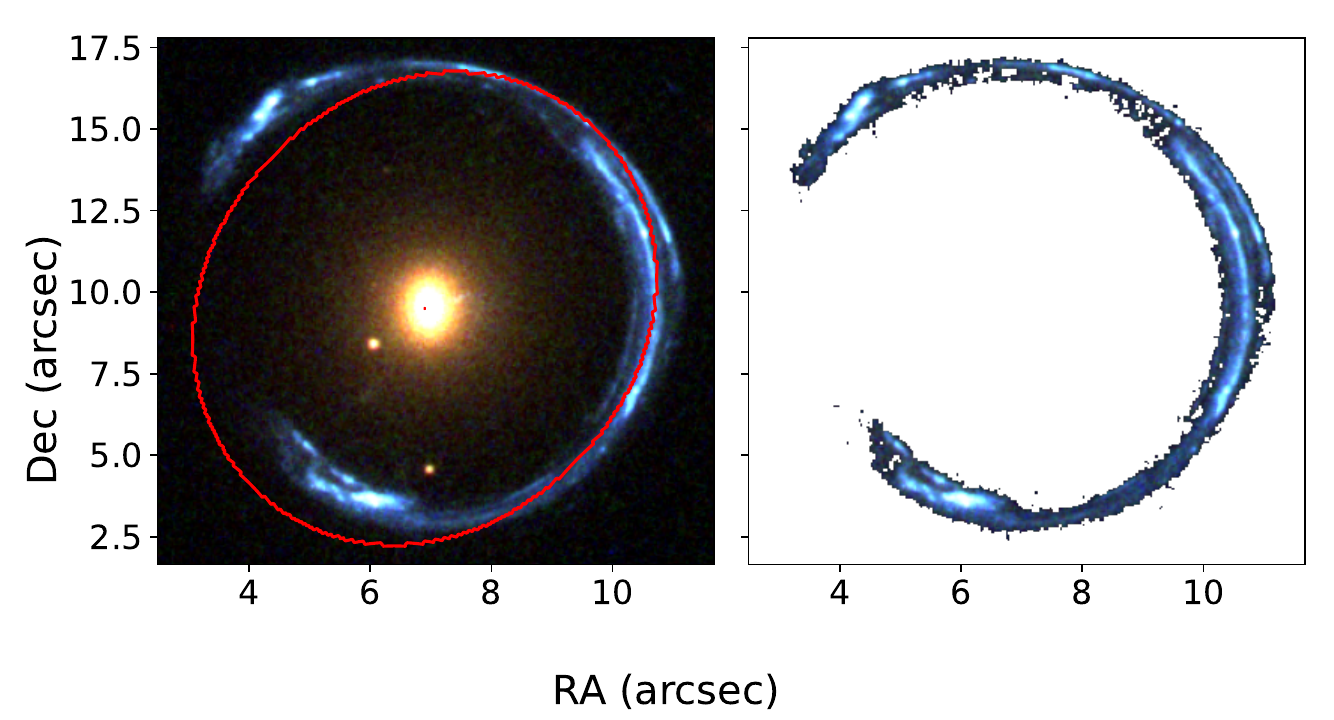}
    \caption{RGB composite image of the Cosmic Horseshoe, featuring F814W in the R channel; F606W in the G channel; and F475W in the B channel. On the left, we show the entire lensing system, overlaid with the critical curve predicted by the \citet{Melo-Carneiro_2025} lens model in red. On the right, we show the same RGB image, but with everything except the arc itself masked out. This is the region where we consider the arc, and predict the transient detection rate upon. The image is oriented North-up and East-right.}
    \label{fig: RGB}
\end{figure}

Here, we study the case of one of the strongest known GGSL systems -- SDSS J114833.14+193003.2, or more commonly known as the ``Cosmic Horseshoe'' \citep[][]{Belokurov_2007}, on whether lensed stars as transients can be detected in this system. At redshift $2.381$ and lensed by a foreground luminous red galaxy at redshift of $0.44$, the Cosmic Horseshoe is one of the largest Einstein rings ever observed (radius of $\sim 5''$), as shown in Figure~\ref{fig: RGB}, with a recent star formation rate estimated between $\sim 50 - 280\,M_{\odot}\textrm{yr}^{-1}$ \citep{Hainline_2009, Quider_2009, Jones_2013, James_2018}. Since transient lensed stars must be intrinsically bright to be detected, the fact that the recent star formation rate in the Cosmic Horseshoe is extremely high leads to an abundance of young, bright stars that might be detectable individually as transients. The Cosmic Horseshoe has a double-source plane with another source forming a radial image at $z = 1.961$, and is also among the most well-modelled GGSL systems \citep[e.g., ][]{Dye_2008, Bellagamba_2017, Cheng_2019, Schuldt_2019}, where recent work has highlighted the use of lens modeling to reveal the mass of the supermassive black hole in the lensing galaxy \citep[][hereafter, M25]{Melo-Carneiro_2025}. The RGB composite image of the Cosmic Horseshoe, alongside the critical curve predicted by the M25 lens model, is shown in the left panel of Figure~\ref{fig: RGB}.


This paper is organized as follows. We first outline the methodology in predicting the transient detection rate in the Cosmic Horseshoe in Section~\ref{sec: method}, for which the corresponding result is shown in Section~\ref{sec: result}. We discuss the benefits of using transients in GGSL in terms of various astrophysical applications in Section~\ref{sec: discuss}.
Throughout this paper, we adopt the AB magnitude system \citep{Oke_1983}, along with standard cosmological parameters: $\Omega_{m} = 0.286$, $\Omega_{\Lambda} = 0.714$, and $H_{0} = 69.6\ \textrm{km s}^{-1}\, \textrm{Mpc}^{-1}$ \citep{Bennett_2014}. The distance modulus at $z = 2.381$ under the adopted cosmological model is 46.46\,mag \citep{Wright_2006}. The critical surface mass density for a lens of $z = 0.44$ and a source of $z = 2.381$ would be 2033\,$M_{\odot}$\,pc$^{-2}$.

\section{Methodology}
\label{sec: method}

In this section, we first describe our methodology in evaluating the stellar luminosity function (sLF), required to compute the number of stars that can be detected as transients, in the Cosmic Horseshoe based on spectral energy distribution (SED) fitting. We then describe the procedure to estimate the transient detection rate based on the inferred sLF.

\subsection{Stellar luminosity function}
\label{sec: sLF}

\begin{figure}
    \centering
    \includegraphics[width=.9\linewidth]{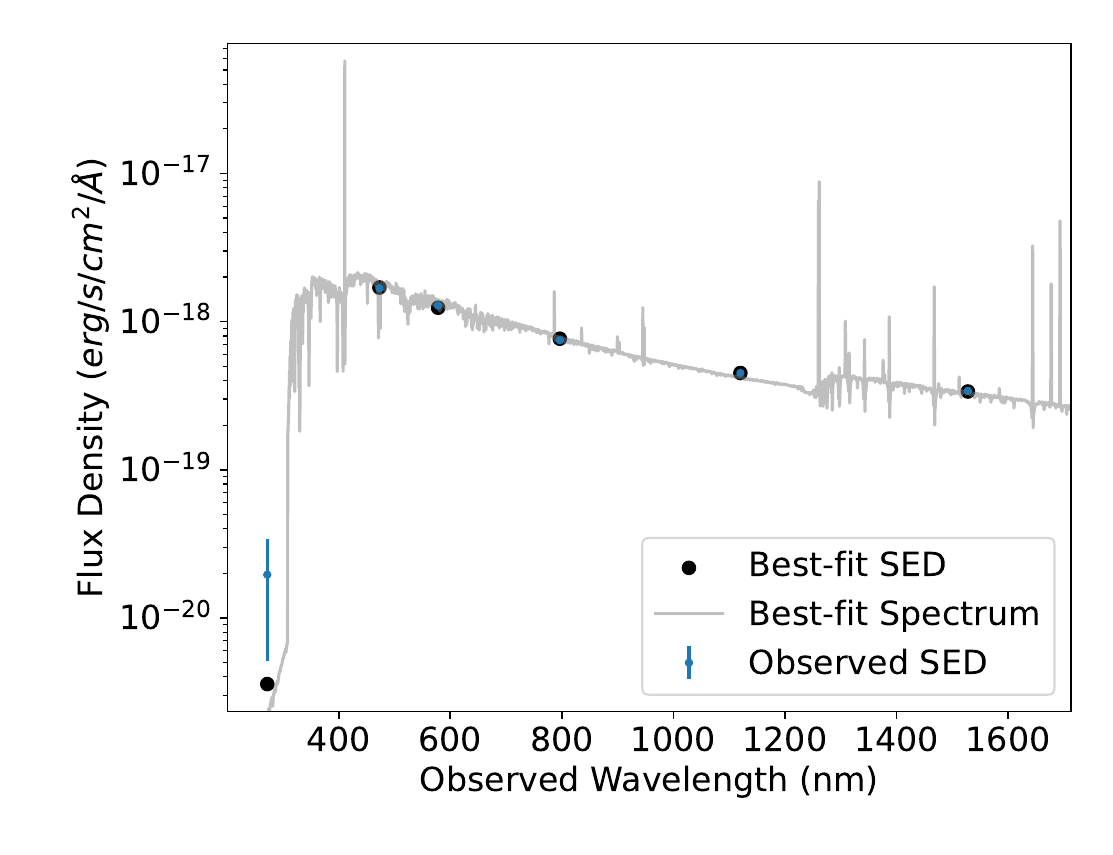}
    \caption{Spectral energy distribution (SED) of the Cosmic Horseshoe, after correcting for lensing magnification as predicted by M25. We used {\tt Bagpipes} to fit a non-parametric star formation history to the observed SED, whereas the best-fitting photometry and best-fitting spectrum are shown as black dots and gray curve, respectively. Notice that the error bars are very small except in F275W, where they are not clearly visible in the SED.} 
    \label{fig: SED}
\end{figure}

We retrieve the {\it Hubble Space Telescope} ({\it HST}) archival images of the Cosmic Horseshoe based on observations under proposals 11602 (PI: S.S. Allam) and 12266 (PI: A.M. Quider). The observations were conducted using the F275W, F475W, F606W, F814W, F110W, and F160W filters. We select regions near the Einstein radius from the F475W image (where the signal-to-noise is the highest) that are brighter than $3\sigma$ of the median sky background as the lensed arc. The mask, applied to the RGB image, is shown in the right panel of Figure~\ref{fig: RGB}.
To remove the light of the lensing galaxy, we first model the lensing galaxy in each of the aforementioned filters (with the arc masked out) as a Sersic profile with IMFIT \citep{imfit}, and subtract it from the images to obtain a clean image for the Cosmic Horseshoe. We then divide the signal of each of the relevant pixels in all these images by the lensing magnification predicted by the M25 lens model. This way, we obtain the delensed SED of the Cosmic Horseshoe, as shown in Figure~\ref{fig: SED}. Notice that we do not account for image multiplicity at this stage.

Since the transient detection rate depends heavily on the most recent star formation history over the last $\sim 50\,$Myr, we also include equivalent width (EW) measurement of four emission lines that are known to correlate with (recent) star formation rate. We do not use the emission line fluxes, as that would depend strongly on the magnification, where EW is lensing invariant.
These four emission lines are H$\alpha$, H$\beta$, $[\textrm{O III}]$ and $[\textrm{O II}]$ respectively. The observed frame EW are extracted from the spectra presented in \citet{Hainline_2009}, for which we find $EW_{H\alpha} = -27.78\pm0.37\,$\AA, $EW_{H\beta} = -10.77\pm0.62\,$\AA, $EW_{\textrm{[O III]}} = -15.00\pm0.12\,$\AA, and $EW_{\textrm{[O II]}} = -21.74\pm0.73\,$\AA, respectively\footnote{Negative sign means it is an emission line.}. Notice that the aperture of \citet{Hainline_2009} is placed on the two brightest regions of the Cosmic Horseshoe at northeast and southwest, for which we anticipate the star formation rate to be the highest. Since the star formation rate is expected to vary in the arc, adopting the emission line EW at the brightest region might have led to an overestimation of the star formation rate of the arc. We shall discuss the effect of having adopted a (possibly) larger EW of the emission lines later in Section~\ref{sec: uncertainty}.

We carried out SED fitting with {\tt Bagpipes} \citep{Bagpipes} to infer the star formation history (SFH) and therefore estimate the number of stars in the Cosmic Horseshoe that could be detected as transients. To provide greatest flexibility in modelling the SFH and thus best capture the variation in SFH, we adopt a step function SFH model (also known as non-parametric model) that allows the stellar mass formed (or the star formation rate over time $t$, $\Psi(t)$) in each age bin, $1-10\,$Myr, $10-50\,$Myr, and $50\,$Myr$-2.5\,$Gyr, to be freely solved by {\tt Bagpipes} \citep{Leja_2019}. The choice of binning is based on minimizing the number of free parameters while obtaining a reasonable fit of the SED, and is discussed further in Appendix~\ref{sec: appendix_SFH}. We shall also discuss later in Section~\ref{sec: discuss} that the choice of SFH model (both the number/choice of binnings in the step-function SFH, or other parametric models) does not significantly affect the inferred transient detection rate.

Ultraviolet spectroscopy reveals photospheric absorption lines and P Cygni profiles (characteristic of stars that have strong stellar winds, such as supergiants, thus indicating very recent star formation) in the Cosmic Horseshoe, for which \citet{Quider_2009} infers a stellar metallicity of $\sim 0.5 Z_{\odot}$. We hence adopt a normal prior in the metallicity inference of the Cosmic Horseshoe, centred at $Z_{\star} = 0.5 Z_{\odot}$ with standard deviation $0.1 Z_{\odot}$, which is approximately the quoted uncertainty in \citet{Quider_2009}. We allow the magnitude of dust extinction measured at V band, $A_{V}$, to be freely varying between $0 < A_{V} < 1$ with a flat prior. 
Given the strong line-emitting nature of the Cosmic Horseshoe, we also allow the strength of nebular emission (in terms of degree of ionization in gas, $\textrm{log}_{10}(U)$) to be freely solved while assuming it has the same metallicity as the stellar population. 
We adopt a \citet{Kroupa_2001} IMF truncated at an upper mass limit of $100\,M_{\odot}$ for the calculation. The best-fitting SED and the underlying spectrum are shown in Figure~\ref{fig: SED}, as black data points and gray curve, respectively. The result of the SED fitting is listed in Table~\ref{tab: SED}. We found a metallicity of $\sim 0.4 Z_{\odot}$ and dust extinction of $A_{V} = 0.6$, similar to the findings in \citet{Hainline_2009, Quider_2009, James_2018}.

\begin{figure}
    \centering
    \includegraphics[width=0.9\linewidth]{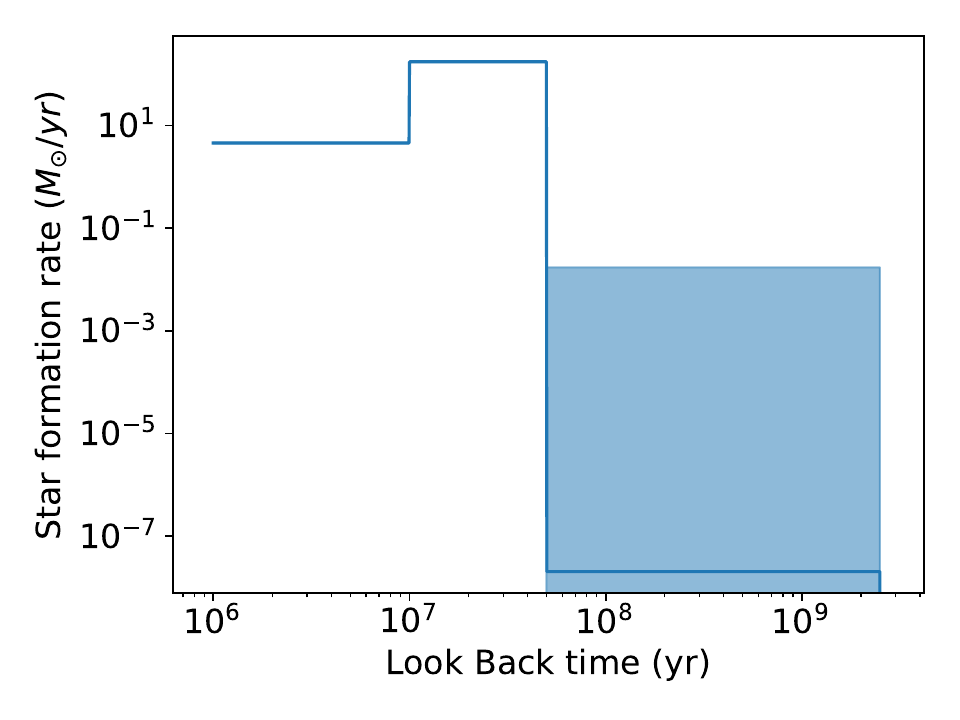}
    \caption{Inferred star formation history for the Cosmic Horseshoe, based on the SED fitting as shown earlier in Figure~\ref{fig: SED}. The mean star formation rate over the last $\sim 50\,$Myr is estimated at $143.4 \pm 2.0 M_{\odot}\textrm{yr}^{-1}$. The shaded regions overlaying the three age bins (which are too small to be shown clearly in the first two bins) represent the $\pm 1\sigma$ uncertainty of the star formation rate inferred from {\tt Bagpipes}'s Markov-Chain Monte-Carlo sampling. The best-fitting parameters and the associated uncertainties are shown in Table~\ref{tab: SED}. }
    \label{fig: SFH}
\end{figure}

\begin{table}
    \centering
    \begin{tabular}{l|c}
     Parameter & Inference \\
     \hline
     Metallicity $,Z_{\star}$    & $0.39\pm 0.02 Z_{\odot}$ \\
     Dust extinction ,$A_{V}$    & $0.61\pm0.12\,$mag \\
     Ionization parameter, $\textrm{log}_{10}(U)$& $-2.90\pm0.01$ \\
     $\Psi(1-10\,\textrm{Myr})$    & $4.59\pm 0.07\, M_{\odot}\textrm{yr}^{-1}$ \\
     $\Psi(10-50\,\textrm{Myr})$     &  $175.10\pm 2.43\, M_{\odot}\textrm{yr}^{-1}$ \\
     $\Psi(50\,\textrm{Myr}- 2.5\,\textrm{Gyr})$     &  $2.06 \times 10^{-8}\pm 0.02\, M_{\odot}\textrm{yr}^{-1}$\\
     \hline
    \end{tabular}
    \caption{Parameters and inferred values from the SED fitting.}
    \label{tab: SED}
\end{table}

\begin{figure}
    \centering
    \includegraphics[width=\linewidth]{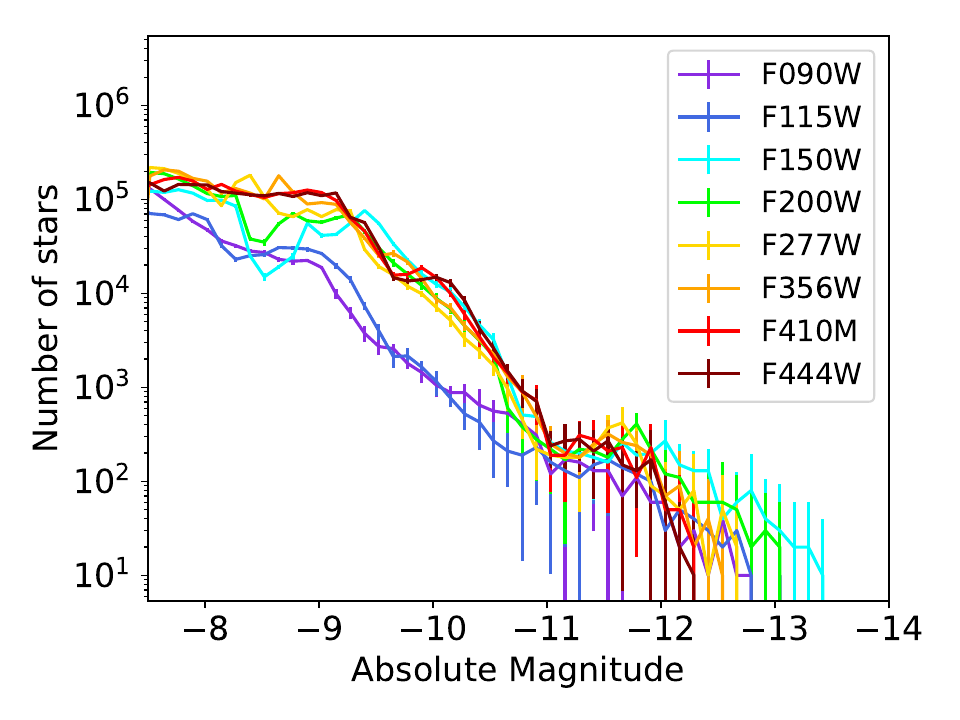}
    \caption{Simulated stellar luminosity function of the Cosmic Horseshoe at eight selected {\it JWST} filters (with different colours) based on the star formation history shown in Figure~\ref{fig: SFH}, as inferred from the SED fitting shown in Figure~\ref{fig: SED}. The uncertainties are propagated from the SED fitting, as well as from the sampling uncertainty during the 30 realizations of stellar population synthesis via {\tt SPISEA}. The fluctuation in the sLF is owing to the intrinsic variation in stellar evolution, as described by the isochrones. We only show those with absolute magnitude brighter than $-7.5$, as stars dimmer than this limit would never be detected, limited by the maximum lensing magnification they can attain. }
    \label{fig: sLF}
\end{figure}

The best-fitting SFH is shown in blue in Figure~\ref{fig: SFH}. The fit has a mean recent star formation rate of $143.4 \pm 2.0\,M_{\odot}\textrm{yr}^{-1}$ over the last $\sim 50\,$Myr, between the literature result of $\sim 50-80\,M_{\odot}\textrm{yr}^{-1}$ from SED UV continuum \citep[which assumed a constant star formation rate over last 100 Myr and do not share the same flexibility as our SFH model,][]{Hainline_2009, Quider_2009}, and $\sim 73-289\,M_{\odot}\textrm{yr}^{-1}$ \citep{Hainline_2009, Jones_2013} inferred from $H\alpha$ luminosity. Since {\tt Bagpipes} adopts a Markov-Chain Monte-Carlo to do the SED fitting, the fact that it prefers a solution with a small uncertainty associated with the recent star formation rate under the full exploration in the parameter space, especially with a step-function SFH that enables the maximum possible flexibility (regardless of the choice of age bins, see Appendix~\ref{sec: appendix_SFH}), indicates that the inferred recent star formation rate should be reliable. The total stellar mass in the Cosmic Horseshoe is $\sim 10^{9.8} M_{\odot}$, in good agreement with the result in \citet{Quider_2009, Jones_2013}.
Based on this SFH, we simulate the background stellar population in the Cosmic Horseshoe with the stellar population synthesis code {\tt SPISEA} \citep{SPISEA} which adopts a MIST isochrone \citep{MIST}. In this way, we compute the sLF at filters $f$, $N(m_{f})$, of the Cosmic Horseshoe. The chosen filters are {\it JWST} F090W, F115W, F150W, F200W, F277W, F356W, F410M, and F444W respectively. The eight sLFs are shown in Figure~\ref{fig: sLF}.

\subsection{Transient detection rate}

Given the sLF of the Cosmic Horseshoe, we can evaluate the transient detection rate in each pixel that represents the arc following the methodology described in \citet{Palencia_2025_microlensing} and \citet{Li_2025_IMF}, which has demonstrated the capability to reproduce the observed transient detection rate in galaxy clusters down to $\sim 1-2 \sigma$ level. 
By combining information on how likely stellar microlensing can further magnify (or demagnify) background stars under the strong lensing effect, we calculate the probability that background stars can be brightened above some detection threshold $m_{5\sigma}$. Such a probability can be computed via the microlensing probability density function, $p$, in \citet{Palencia_2024}, which characterize the probability based on: (1) the tangential magnification, $\mu_{t}$, (2) the radial magnification, $\mu_{r}$, and (3) abundance of stellar microlenses, $\Sigma_{\star} (M_{\odot}/\textrm{pc}^{2})$:

\begin{equation}
    p(\mu; \mu_{t}, \mu_{r}, \Sigma_{\star})
\end{equation}

\begin{figure*}
    \centering
    \includegraphics[width=\linewidth]{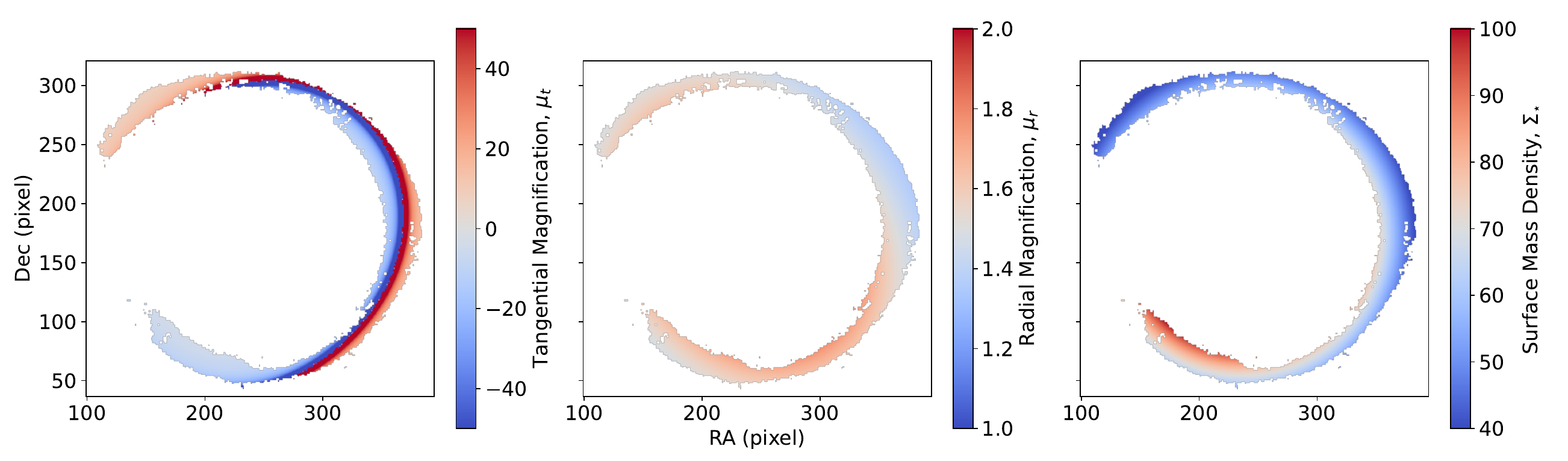}
    \caption{Predicted tangential magnification (left), radial magnification (middle), and stellar surface mass density (right) from the M25 lens model. }
    \label{fig: lens_model}
\end{figure*}

\noindent We again adopt the M25 lens model to acquire $\mu_{t}$, $\mu_{r}$ and $\Sigma_{\star}$ for each pixel that represent the Cosmic Horseshoe. The distribution of all these three variables across the arc is shown in Figure~\ref{fig: lens_model}.

As we are interested in transients, this means that the background stars are not detectable without the extra magnification from microlensing. In other words, the stars that can be detected as transients must have apparent magnitude in filter $f$, $m_{f}$, dimmer than the detection threshold $m_{5\sigma, f}$ when there is only macro magnification $\mu_{m} = \mu_{t}\mu_{r}$. Mathematically:

\begin{equation}
    m_{f} - 2.5 \textrm{log}_{10}(\mu_m) \geq m_{5\sigma, f} \,.
\end{equation}

\noindent Such a limit is referred to as the Detectable-Through-Microlensing limit \citep{Diego_2024_3M}: stars that satisfy such a condition are dubbed as DTM stars. 

Previously, we have estimated the sLF of the Cosmic Horseshoe. Different image plane pixels having different brightnesses contain a different number of stars in the source plane. To calculate the number of stars in each pixel, we distribute the total sLF of each $i$ pixel following their intrinsic brightness (i.e., corrected by macro magnification) in F475W ($F_{i, F475W}$) compared with the intrinsic brightness of the entire arc in F475W\footnote{This would have assumed that the photometric colour of the arc is homogenous thus the underlying sLFs are invariant, which is very much the case when one looks at Figure~\ref{fig: RGB}.}, the filter where the arc is the brightest with the highest signal-to-noise:

\begin{equation}
\label{eqn: scale_flux}
    N_{i}(m_{f}) = N(m_{f}) \times \frac{F_{i, F475W}/\mu_{m, i}}{\sum^{N}_{i} F_{i, F475W}/\mu_{m, i}}\, .
\end{equation}

\noindent In this process, the image multiplicity is automatically accounted for. Then, for each pixel $i$, we can calculate the sLF at filter $f$ after correcting for lensing magnification as, $N_{i}(m'_{f})$:

\begin{equation}
\label{eqn: integral}
   N_{i}(m'_{f}) = \int_{\mu_{\rm min}}^{\mu_{\rm max}}d\mu  [p(\mu; \mu_{t, i}, \mu_{r, i}, \Sigma_{\star, i}) N_{i}(m_{f}-2.5\textrm{log}_{10}\mu)]\,  ,
\end{equation}

\noindent $\mu_{\rm min}$ and $\mu_{\rm max}$ depend on the simulation resolution and source size, respectively. We adopt $\mu_{\rm min} = 10^{-1}$ following the original work in \citet{Palencia_2024}. The choice of $\mu_{\rm min}$ does not matter -- so long as it is well below the level where a transient can result, as it would not affect the detection probability. The $\mu_{\rm max}$ depends on the source radius of DTM stars ($R_{\star}$), the mass of the microlens (thus their Einstein radius, $\theta_{E}$), as well as the strong lensing effect ($\mu_{t}$ and $\mu_{r}$). We adopt the approximation in \citet{Oguri_2018} that $\mu_{\rm max} \approx \mu_{t}^{3/4}\mu_{r}\sqrt{\theta_{E}/R_{\star}}$. The typical mass of microlenses is assumed to be $\sim 0.3 M_{\odot}$ \citep{Meena_2025_IMF}, and the typical radius would be $\sim \mathcal{O}(100) R_{\odot}$ for red DTM stars (red supergiants), and $\sim \mathcal{O}(10)R_{\odot}$ for blue DTM stars (blue supergiants). The $\mu_{\rm max}$ is thus filter and pixel dependent, roughly given as $\sim \sqrt{300} \times \mu_{r}\mu_{t}^{3/4}$ for red DTM stars (filters F150W to F444W, restframe optical to near-infrared), or $\sim \sqrt{3000} \times \mu_{r}\mu_{t}^{3/4}$ for blue DTM stars (filters F090W and F115W, restframe UV). With Equation~\ref{eqn: integral}, the detection rate of transients at any given pixel can be obtained by integrating the magnification corrected sLF and finding the expected number of stars that are brighter than the detection threshold:

\begin{equation}
\label{eqn: detection_rate}
    R_{i, f} = \int^{m_{5\sigma, f}}_{-\infty} N_{i}(m'_{f}) dm'\, .
\end{equation}

\noindent Of course, the total detection rate in the whole Cosmic Horseshoe is just the sum of the detection rates of all the pixels that represent the arc itself:

\begin{equation}
\label{eqn: sum_rate}
    R_{f} = \sum^{N}_{i} R_{i, f}\, .
\end{equation}

\section{Result}
\label{sec: result}

\begin{figure}
    \centering
    \includegraphics[width=1\linewidth]{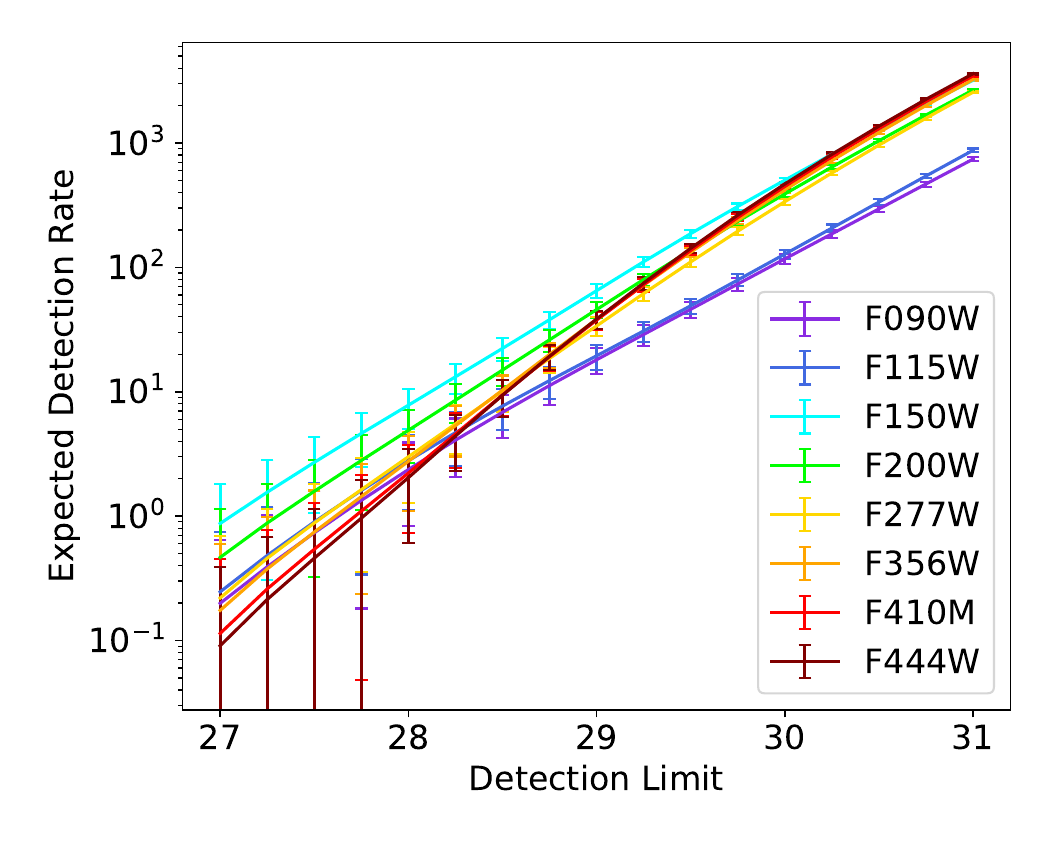}
    \caption{Predicted transient detection in the Cosmic Horseshoe in eight JWST filters, as a function of detection limit in $m_{AB}$. One can see that the detection rate depends on the filter and increases exponentially with a deeper detection limit. The filter with the highest detection rate is F150W, reaching $\sim 1$ at $\sim 27\,m_{AB}$, $\sim 10$ at $\sim 28\,m_{AB}$, and $\sim 60$ at $\sim 29\,m_{AB}$. }
    \label{fig: rate_vs_limit}
\end{figure}

From Equation~\ref{eqn: sum_rate}, we predict the transient detection rate in the Cosmic Horseshoe for the eight selected filters mentioned earlier in Section~\ref{sec: sLF} at different detection limits. The results are shown in Figure~\ref{fig: rate_vs_limit}, where the uncertainty propagates from the uncertainty in the inferred SFH, the sampling uncertainty in generating the sLFs, and the Poisson noise in the detection rate itself. One can see that the detection rate increases exponentially as the detection limit becomes deeper. The exact shape of how the detection rate increases with detection limit is filter-dependent, as the shape of the sLF at different filters is coupled with the stellar evolution. The transient detection is of the order of $\sim 1-10$ per pointing at observations with detection limits $\gtrsim 28\,m_{AB}$, meaning that detecting transients is very possible in the Cosmic Horseshoe with depth similar to {\it JWST} cluster observations. For observations reaching $\sim 28.6\,m_{AB}$, the detection rate is $\sim 25$, comparable with the record holder of transient detection rate -- ``Dragon'' arc \citep[$z = 0.73$,][]{Fudamoto_2025} at the same detection limit.

\begin{figure}
    \centering
    \includegraphics[width=\linewidth]{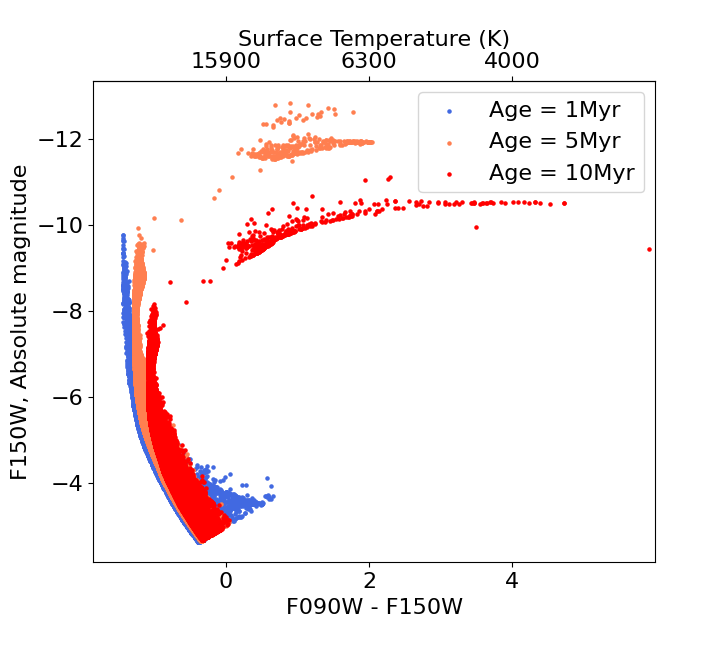}
    \caption{Simulated stellar population for ages of $\sim$1\,Myr (blue), $\sim$5\,Myr (orange), and $\sim$10\,Myr (red) based on the star formation rate shown in Figure~\ref{fig: SFH}, respectively, shown as a colour-magnitude diagram with brightness in F150W versus the colour of F090W-F150W (and surface temperature, assuming a perfect blackbody). One can see that massive stars begin to evolve into red supergiants at an age of $\sim 5\,$Myr, and are always brighter than blue supergiants in either F090W or F150W. This eventually leads to the transient detection rate being the highest at F150W.}
    \label{fig: HR_diagram}
\end{figure}

The detection rate is almost always highest for F150W, which covers a restframe wavelength of $\sim 440\,$nm. This filter corresponds to the wavelength at which a blackbody at a temperature of $\sim 7000\,$K peaks in intensity for a redshift of $2.381$. The brightest class of DTM stars (thus most likely to be detected as transients) with such a characteristic surface temperature would be the red supergiants. While it might be counterintuitive that a very blue stellar population like the Cosmic Horseshoe would host a lot of red supergiants, this can be explained by looking at the SFH shown in Figure~\ref{fig: SFH} -- the high star formation rate sustained for the first $50\,$Myr, where red supergiants begin to appear after $\sim 5\,$Myr as shown in Figure~\ref{fig: HR_diagram}. Notice in Figure~\ref{fig: sLF} that there are more blue supergiants (bright in F090W) than red supergiants (brighter in F150W and filters redward). 
However, the brightest stars are always more likely to be red supergiants than blue supergiants after such a characteristic age, as shown in both Figure~\ref{fig: sLF} and Figure~\ref{fig: HR_diagram}. This explains the boost in detection rate in the F150W filter, compared with that in the F090W filter, as the latter is more sensitive to the blue supergiants. 

The main driver behind such a high detection rate is the high recent star formation rate. At redshift $\sim 2.4$, we are witnessing the growth of an extremely massive galaxy with a stellar mass of $\sim 10^{10}M_{\odot}$ \citep[which aligns with the calculation in ][]{Quider_2009, Jones_2013}. 
Nearly 90\% of the star formation of the Cosmic Horseshoe happened in the last $\sim50\,$Myr, the age in which young, massive DTM stars are forming. This, combined with the fact that the Cosmic Horseshoe is among the largest lensed galaxies on the sky ($\sim 21\,{\rm arcsec^{2}}$, based on our mask of the Horseshoe shown in Figure~\ref{fig: RGB}) and hence has a significant cross-sectional area with the critical curve, eventually leads to an extremely high transient detection rate. We can contrast our result with well-studied lensed galaxies that have transient detection with {\it JWST} observations, for example, ``Warhol'' ($z \approx 1$, lensed by the cluster MACSJ0416), to check whether the expected detection rate is reasonable. Warhol has a detection rate of $\sim 2$ per pointing with F150W, given a detection limit of $29.5\, m_{AB}$ \citep{Williams_2025}. \citet{Palencia_2025_microlensing} deduced a recent star formation rate of $\sim 4\times 10^{-4}M_{\odot}\textrm{yr}^{-1}$ (corrected by lensing magnification), with a total stellar mass of $\sim 10^{6} M_{\odot}$ in Warhol. In contrast, the Cosmic Horseshoe has a star formation rate $5-6$ orders of magnitude higher (depending on the exact period of look back time), and is four orders of magnitude more massive than Warhol. Also, the Cosmic Horseshoe has an image plane area $\sim 4$ times larger than Warhol ($\sim 5\, {\rm arcsec}^{2}$). Our prediction of transient detection rate at F277W (roughly the same rest-frame wavelength as F150W in Warhol) is $\sim 110\pm10$ -- about two orders of magnitude more than that of Warhol. With a $\sim2.5\,$mag dimmer distance modulus (such that stars appear to be $\sim10$ times dimmer) for the Cosmic Horseshoe compared with that of Warhol, our estimated detection rate here appears to be reasonable.

\begin{figure}
    \centering
    \includegraphics[width=\linewidth]{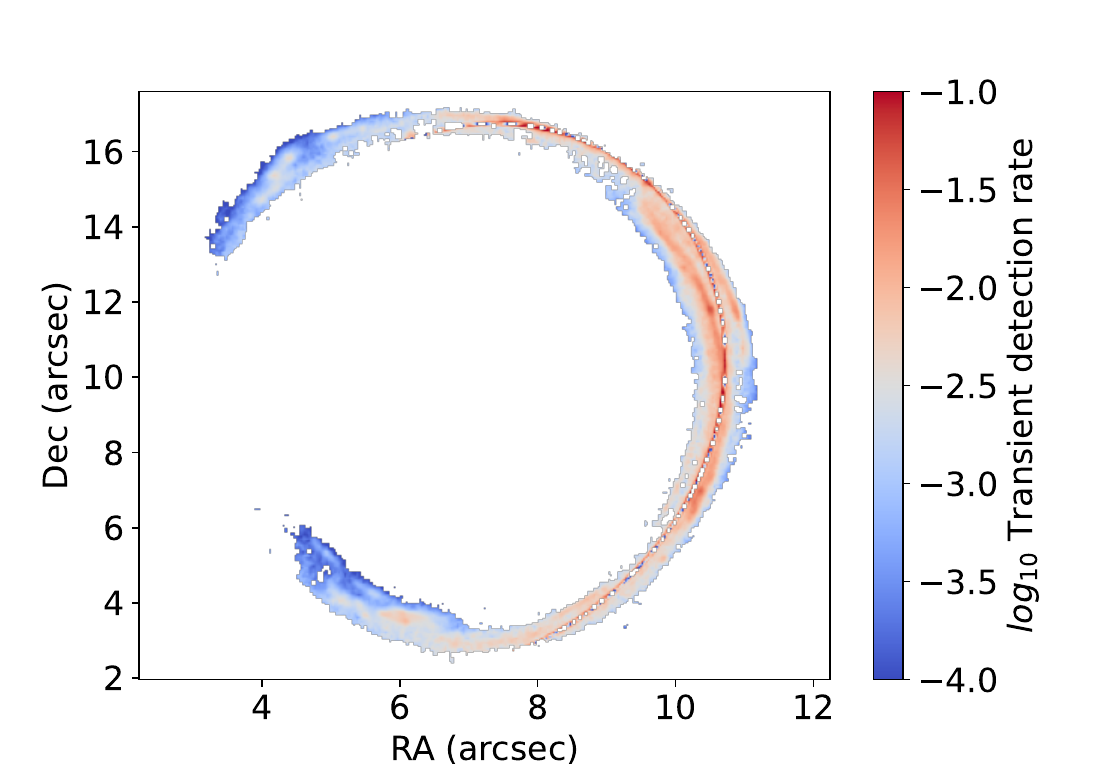}
    \caption{Expected transient detection rate per pointing distributed over the Cosmic Horseshoe, at F150W and a $5\sigma$ detection limit of 29 $m_{AB}$. One can see that the transient detection rate is the highest at the East of the Cosmic Horseshoe, where the critical curve cuts through the arc. }
    \label{fig: rate_map}
\end{figure}

We also show the spatial distribution of the expected detection rate in the Cosmic Horseshoe for F150W with a $5\sigma$ detection limit of 29$m_{AB}$ in Figure~\ref{fig: rate_map}. One can see that the transients are most likely to be detected towards the East direction, where the cross-sectional area of the arc and the high magnification region associated with the critical curve maximizes. This aligns with one's expectation where transients tend to be found near regions with higher macro-magnification \citep[but not the highest, ][]{Palencia_2025_microlensing, Li_2025_IMF}.

\section{Discussion}
\label{sec: discuss}

\subsection{Applications}
\label{sec: applicatoin}

With the prediction of being able to detect transients in GGSLs, it would be interesting to repeat what had been done on lensed stars in clusters on GGSLs. A clear benefit of GGSL systems over clusters is the accuracy of the lens model, as galaxies serving as the main deflector could often be well modeled, compared with the more intricate cluster environment.

\subsubsection{Implication for the nature of dark matter}

Using the spatial distribution of $\sim$40 transients reported by \citet{Kelly_2022_Flashlights} and \citet{Fudamoto_2025}, \citet{Broadhurst_2025} tested predictions of dark matter that could be composed of weakly interacting massive particles and ultra-light axions (also known as wave-like dark matter), respectively. For weakly interacting particles, a positive skewness in the spatial distribution (towards the exterior of the critical curve) is expected \citep[see also][]{Williams_2024}, while a negative skewness (towards the interior of the critical curve) is expected for the wave-like dark matter and better matches the observation. Since the result depends on the spatial distribution of the positions of the transients with respect to the critical curve, it was highlighted in \citet{Broadhurst_2025} that one of the perhaps major uncertainties is the position of the critical curve. Although it was argued via lensing symmetry in several lensed arcs in clusters that the critical curves could be stringently defined down to the pixel level \citep[e.g.,][]{Dai_2020}, the many successes in GGSL lens reconstruction \citep[e.g.,][]{Vegetti_2010, Minor_2021, He_2024, Enzi_2025, Lange_2025} have demonstrated that there is less room for uncertainty in the position of the critical curve in GGSL compared with the cluster lensing case in terms of image reconstruction.

Moreover, the characteristic de Broglie wavelength of wave-like dark matter scales linearly with the axion mass and inversely with the mass of the main halo \citep[$\lambda_{db} \propto m_{\psi}^{-1} M^{-1/3}$, ][]{Schive_2014}. It is an order of magnitude larger in GGSLs (typical mass of $\sim 10^{10-12}\,M_{\odot}$) compared with that in clusters (typical mass of $\sim 10^{14-15}\,M_{\odot}$) given the same axion mass. The immediate consequence is that the effect of wave-like dark matter in broadening the spatial distribution of transients along the critical curve should be more pronounced in GGSLs \citep[see also discussion in ][]{Amruth_2023, Palencia_2025_waveDM}. \citet{Broadhurst_2025} predicted how the width (specifically, full-width at the half-maximum since the distribution is Gaussian-like) of the spatial distribution of transients across the critical curve scales with the Einstein radius. Following the same correlation, we calculate the width as $\sim 0.0044\, \theta_{E} \sqrt{\lambda_{db}/1\textrm{pc}}$ for the Cosmic Horseshoe, which has a total projected mass of $5.5\times10^{12}M_{\odot}$ within the Einstein radius \citep{Melo-Carneiro_2025}. Given that the Einstein radius of the Cosmic Horseshoe is $\sim 5''$ \citep{Belokurov_2007}, the expected width would be $\sim 0.3''$, equivalent to $\sim 1.6\,$kpc, if we assume a typical boson mass of $\sim 10^{-22}\,\textrm{eV}$ \citep{Schive_2014, Hui_2021}. With the apparent thickness of $\sim 0.5-1''$ of the Cosmic Horseshoe, the broadening of the distribution of transients due to wave-like dark matter could be well captured within the arc, and is more appreciable than the case (in terms of broadening versus the width of the arc) in \citet{Broadhurst_2025}. This, adding to the fact that there is less room for uncertainty in the position of the critical curve, suggests that GGSL systems like the Cosmic Horseshoe could be an excellent testing ground for examining whether dark matter is constituted of ultra-light axions or weakly interacting massive particles, complementary to galaxy cluster observations. Notice that this benefit is independent of the transient detection rate, as it only affects the number of observations required to have a sufficient amount of samples to distinguish between different classes of dark matter particles, or to constrain the dark matter mass.



\begin{figure}
    \centering
    \includegraphics[width=1\linewidth]{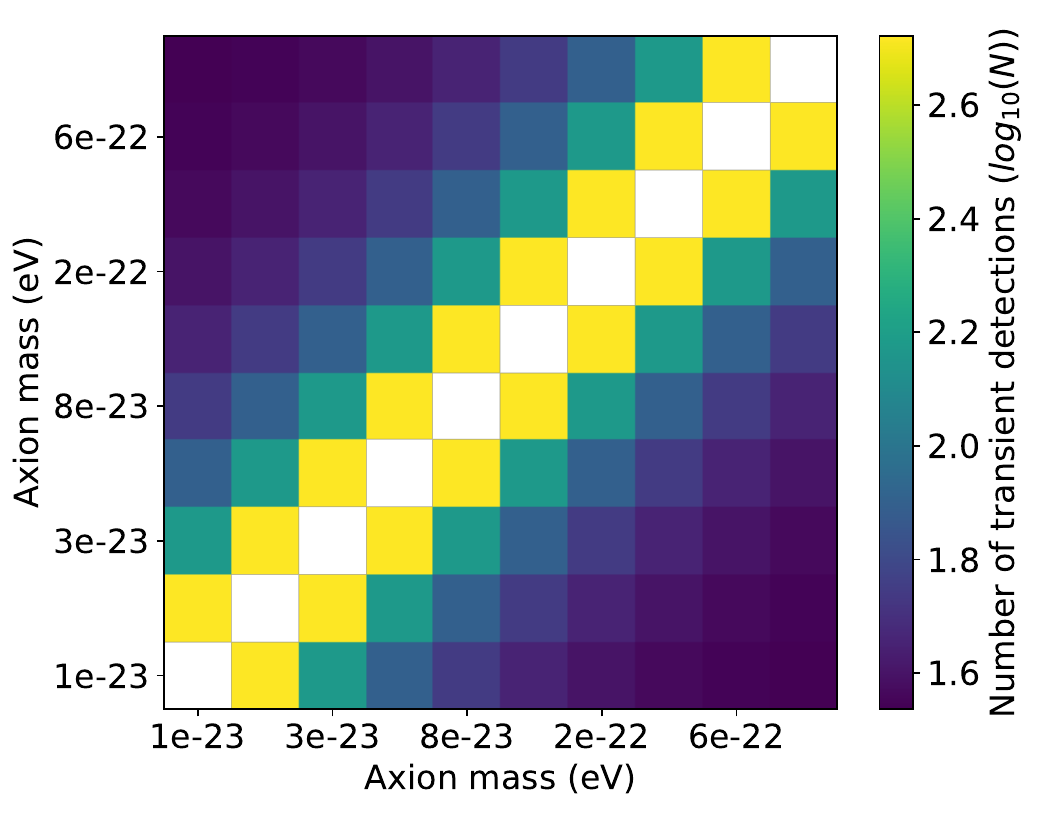}
    \caption{Number of transient detections (as indicated by pixel colours) that are required to distinguish any given two axion masses (x and y axes) down to $3\sigma$ confidence level. The diagonal entry is null as one can never tell the same axion mass apart. }
    \label{fig: m_psi}
\end{figure}

We can perform a simple calculation to estimate how many observations on the Cosmic Horseshoe are required to tell different axion masses apart if dark matter is constituted of ultra-light axions, given our predicted transient detection rate. Since we are constraining the axion mass based on the width of the distribution, we can utilize the F-test to compare the variances of two Gaussian distributions and estimate how many observations on the Cosmic Horseshoe are required to tell two axion masses apart at $3\sigma$ confidence. We show the number of transients required in Figure~\ref{fig: m_psi}. Both axes show the Axion mass, where the colour of the image shows the number of transients that are required to be detected to separate the two Axion masses in the two axes.

The number of observations required (which, of course, has to be greater than two since one cannot observe a transient with only one epoch) is just the number of transients that are required to be detected, divided by the expected transient detection rate given some detection limit. Since the statistics here are not colour sensitive, we can always choose the band with the highest detection rate, i.e., F150W, as mentioned in Section~\ref{sec: result} to carry this exercise on. With a detection limit of $\sim 29\,m_{AB}$, we would expect $\sim 60$ transients per pointing in F150W. The number of transients that is required to tell Axion masses of $10^{-23}\,$eV and $10^{-21}\,$eV apart is just $\sim 30$ (lower right pixel in Figure~\ref{fig: m_psi} -- meaning that we can already distinguish between the two Axion masses in just two observations at $29\,m_{AB}$. Of course, one would need more observations to tell two more similar axion masses apart. For example, we would need $\sim 10$ observations ($\sim 500$ transients) at the same detection limit to separate axion masses of $8\times10^{-23}\,$eV and $1\times10^{-22}\,$eV apart with $3\sigma$ confidence. As a reference, observations with galaxy clusters that have masses $\sim 3$ orders of magnitude more than galaxy lenses would require $\sim \sqrt{10} \approx 3.3$ times more observations to constrain axion mass as the de Broglie wavelength is $\sim (10^{3})^{-1/3} \approx 10^{-1}$ times shorter. This highlights the benefits of using the spatial distribution of transients in GGSL systems to constrain the axion mass over galaxy cluster lenses.

\subsubsection{Initial mass function at the cosmic noon}


The current record holder for an arc that has discovered the most transients is the ``Dragon'' \citep{Kelly_2022_Flashlights, Fudamoto_2025}, which resides at a rather low redshift of $\sim 0.73$. The Cosmic Horseshoe is at a redshift of $2.38$, looking further back $\sim 4.4\,$Gyr in time ($\sim 2.8\,$Gyr since the Big Bang) compared with the Dragon. Our predicted, comparable transient detection at the Cosmic Horseshoe makes it an ideal target to observe individual stars at the cosmic peak of star formation between redshifts 2-3 \citep{madau_2014}. 


\begin{figure*}
    \centering
    \includegraphics[width=\linewidth]{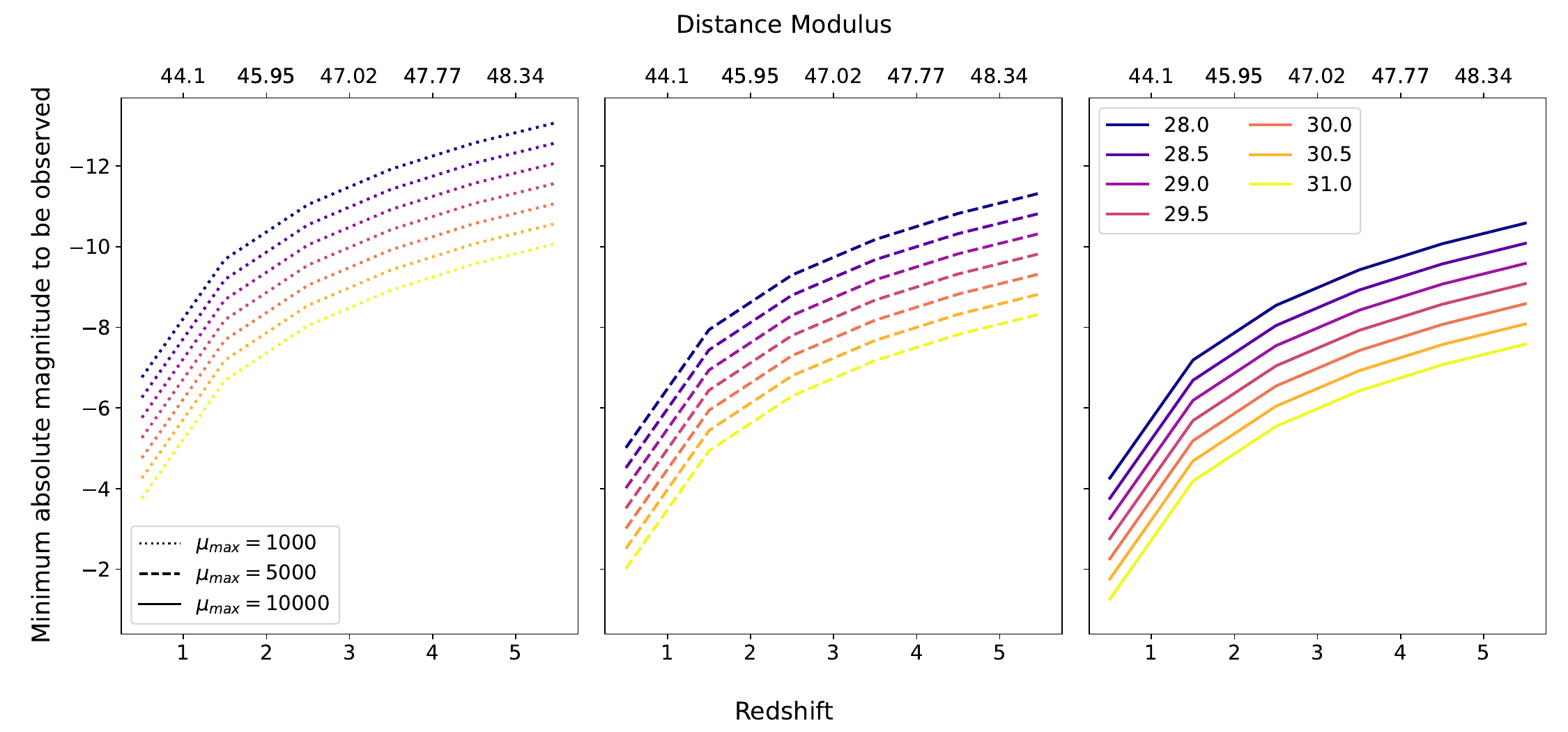}
    \caption{Minimum absolute magnitude for a background star to be detected as a transient as a function of redshift (and distance modulus) given different detection limits (represented by the colour, refer to the legend in the right panel). We show the three cases where the maximum magnification (which is inversely correlated to the radius of the star, given the same macro-magnification) that the background star can attain is 1000 ($\equiv 7.5\,$mag), 5000 ($\equiv 9.2\,$mag), and 10000 ($\equiv 10\,$mag) in the three columns, respectively. One can see that the redshift (thus the distance modulus) can be seen as a filter of stellar population, where one can only see more and more massive stars with increasing redshift, given the same detection limit. This increases the sensitivity of the detection rate towards the stellar IMF.}
    \label{fig: DM_filter}
\end{figure*}

In usual circumstances, the increased redshift (thus distance modulus) is a challenge as it limits the minimum brightness (thus mass) of stars that we can observe via strong lensing and microlensing combined, hence decreasing the detection rate in general terms. We show the minimum absolute magnitude of background star that is required to detect under different detection thresholds, at given maximum magnification attainable \citep[which is inversely correlated with the stellar radii, ][]{Oguri_2018} as a function of redshift (thus, distance modulus) in Figure~\ref{fig: DM_filter}. One can see that transient detection is limited to intrinsically more luminous and more massive stars with increasing redshift. For $z = 0.73$, the distance modulus is 43.3, $\sim 3\,$mag brighter than that at $z = 2.4$. This means that given the same maximum attainable magnification and the same detection threshold, one would be seeing $3\,$ magnitude shallower in the sLF when looking at lensed stars at $z = 2.4$. For the case of the Cosmic Horseshoe, the decrease in detection probability owing to the increased distance modulus is overcome by the extremely high star formation rate. Since the detection rate is not a problem for the Cosmic Horseshoe, the increase in the distance modulus can be conversely viewed as a filter to screen out less massive stars. This is because at lower redshift, intrinsically dimmer stars can have high lensing magnification and have the same apparent brightness as intrinsically brighter stars that have lower lensing magnification -- the intrinsic brightness and the magnification attained is degenerate. However, given the same detection limit, only the intrinsically brighter (thus more massive) stars can be detected at higher redshift as the maximum magnification is limited. In other words, we can obtain a purer sample of the most massive, thus luminous stars at the cosmic noon, providing an even more direct, sensitive view towards the most massive stars formed in the most recent episodes of star formation.

Given our predicted transient detection rate inferred from a SED-consistent SFH, repeated observation carried out on the Cosmic Horseshoe would generate a large number of statistics to infer stellar properties (e.g., stellar IMF) associated with massive stars at $z \approx 2.4$. For example, it was measured in two $z \approx 1$ lensed galaxies via the transient detection rate that they do not have an IMF with high-mass ($M \geq 1.4M_{\odot}$) power law slope that is significantly different from the locally measured value of $\alpha \approx 2.3$  \citep{Li_2025_IMF, Williams_2025}. That said, \citet{Meena_2025_IMF} marginalized the transient detection of $\sim 10$ lensed arcs across six galaxy clusters over different redshifts and suspected that the slope of the IMF might have become shallower than the locally measured value at redshifts between $\sim 1$ and $\sim 3$. The apparent downside of this study is the low transient detection rate at $z \gtrsim 2$ which could cause the inference to be biased towards the lower redshift arcs. Thus, the Cosmic Horseshoe could be a perfect complementary target to supply transient detections at the cosmic noon. The especially high sensitivity of detection rate in the Cosmic Horseshoe towards the massive stars (and therefore the slope of the IMF) compared with lower redshift star-forming galaxies \citep[such as ``Spock'', $z = 1.0$; and ``Warhol'', $z = 0.94$, ][]{Li_2025_IMF, Williams_2025}, combining with the high predicted transient detection rate ($\gtrsim 50$ at $5\sigma$ detection limit of $\sim 29\,m_{AB}$) would allow us to test a variety of slopes of the IMFs.

\begin{figure}
    \centering
    \includegraphics[width=\linewidth]{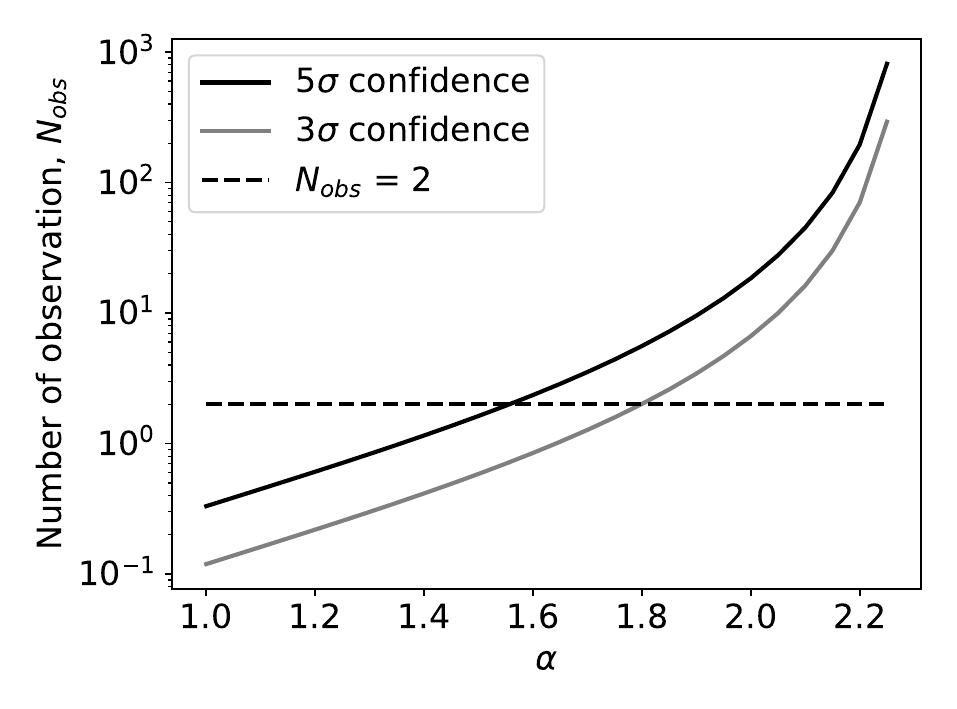}
    \caption{Number of observations that is required to distinguish a given stellar IMF with power law slope $\alpha$ (x-axis) from the locally measured $\alpha = 2.3$ with $5\sigma$ confidence (black solid line) and $3\sigma$ confidence (grey solid line). We assume a $5\sigma$ detection limit of $29\,m_{AB}$ at {\it JWST} F150W -- the filter which has the highest detection rate as shown in Figure~\ref{fig: rate_vs_limit}. The dashed horizontal line shows the level of two observations -- the minimum number of observations required to detect a transient. }
    \label{fig: IMFs}
\end{figure}

To test whether this is feasible, we predict the SED-consistent transient detection rate again as described in Section~\ref{sec: method}, but with a top-heavy IMF ($\alpha = 1$ for all stellar masses) instead of a Kroupa IMF ($\alpha = 2.3$ for stars with masses $\gtrsim 1.4 M_{\odot}$). The transient detection rate increases drastically with a shallower IMF, as there exist more massive, thus more luminous stars in the lensed galaxy \citep{Li_2025_IMF, Meena_2025_IMF, Williams_2025}. Given the same detection limit of $29\,m_{AB}$ at filter F150W, the Cosmic Horseshoe with a top-heavy IMF would have $\sim 200$ transients per pointing, $\sim 3$ times more than the prediction for the case of a Kroupa IMF. Since the simulation uncertainty (which includes uncertainty in the SED fitting) is dominated by Poisson noise and is rather small, two observations alone can distinguish whether the Cosmic Horseshoe has a top-heavy IMF or a Kroupa IMF. Of course, we have been testing two extreme IMFs, and more observations would be required to distinguish between less distinctive IMFs. If we simply interpolate the detection rate $R_{\textrm{F150W}}$ at detection limit of $29\, m_{AB}$ from our simulation, as a function of $\alpha$ and assume they correlate as a power law (as the IMF is also a power law), we can estimate the number of observations required to distinguish between any given slope of IMFs with the locally observed $\alpha \approx 2.3$, to the first-order, as shown in Figure~\ref{fig: IMFs}. As the minimum number of observations required to detect a transient is two (dashed horizontal line in Figure~\ref{fig: IMFs}), two observations on the Cosmic Horseshoe are enough to distinguish between $\alpha \lesssim 1.5$ and $\alpha = 2.3$ down to $5\sigma$ confidence, or between $\alpha \lesssim 1.8$ and $\alpha = 2.3$ with $3\sigma$ confidence. For $\alpha$s that are more similar to $\alpha = 2.3$, one would need more observations to distinguish between them. Since we only aim to deliver an event rate prediction in this paper, we defer a thorough analysis on the dependency of the constraining power of IMF on the adopted SFH model for future works, especially when {\it JWST} observations are available, which would allow one to better constrain the SFH.

\subsection{Archival search of lensed stars}


One interesting point to consider is whether we have already seen some lensed stars. Although transients are only detected by comparing two epochs of observations, they could have been detected in one observation, and we only do not know that they will fade away later. These lensed stars could have then been misidentified as regular compact objects, such as lensed star clusters. As we have predicted earlier in Section~\ref{sec: result}, the detection rate of transients is the highest in {\it JWST} F150W, which has a similar throughput compared with {\it HST} F160W, where the Cosmic Horseshoe is observed. The $5\sigma$ detection limit in the F160W image is $\sim 27.3$ as calculated from recovering injected point sources via photometry, as shown in Appendix~\ref{sec: appendix_limit}. According to our prediction shown earlier in Figure~\ref{fig: rate_vs_limit}, the expected transient detection rate in F150W is $\sim 1.5_{-1.2}^{+1.3}$ at a $5\sigma$ limit of 27.3. In other words, it is not impossible that we have already detected a lensed star in the sole epoch of F160W.

\begin{figure*}
    \centering
    \includegraphics[width=\linewidth]{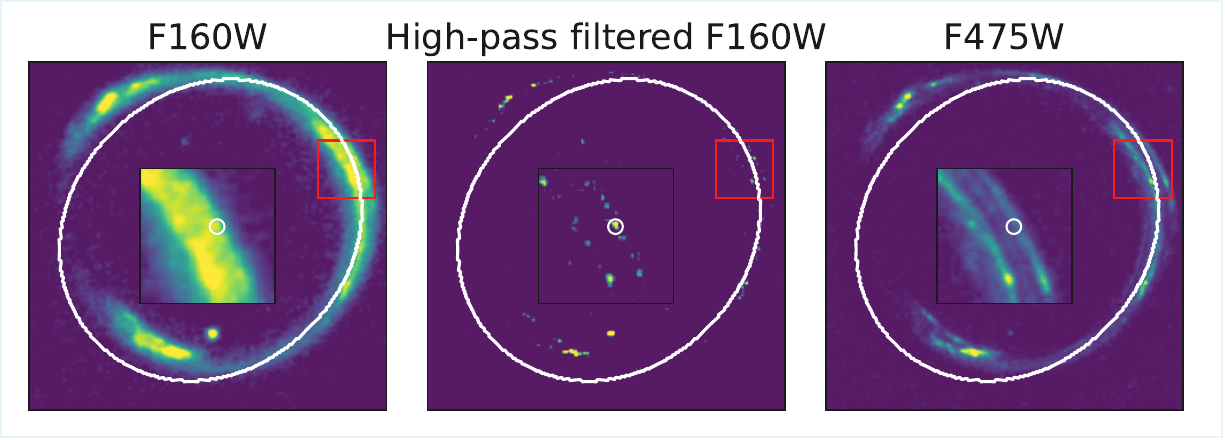}
    \caption{High-pass filtered {\it HST} F160W image of the Cosmic Horseshoe (middle panel), with the original F160W image shown in the left panel and the F475W image (which has the highest signal-to-noise, and is used for lens reconstruction in M25) shown in the right. The critical curve of the M25 lens model is shown as white in all three panels. 
    A small signal is picked up in the filtered F160W image, as circled by the white aperture in the zoomed-in inset figure in the middle. The region of zooming in is denoted by the red square in the original FOV. This signal has no correspondence in the F475W image, let alone being reproduced in the lens reconstruction. However, the photometry reveals that this detection is $\lesssim 3\sigma$, so it is not a significant detection.
    }
    \label{fig: filtered_image}
\end{figure*}

To search for compact features that could possibly be lensed stars, we apply a high-pass filter on the F160W image that has the lensing galaxy subtracted away, as shown in the left panel of Figure~\ref{fig: filtered_image}. We first convolve the image with a 2D Gaussian Kernel with a full-width at half-maximum of 3 pixels \citep[$0.12''$, ][]{Diego_2024_MACSJ0416_lensmodel}, then subtract this convolved image from the original, lensing galaxy-subtracted image as shown in the middle panel of Figure~\ref{fig: filtered_image}. The lower limit of the image scale is set at three times the root-mean-square noise of the high-pass image, such that any signal below the $3\sigma$ rms noise level would not appear on the image. 


We search for compact features revealed in the high-pass image. By so doing, we spotted one possible transient candidate as circled in white in the inset images of Figure~\ref{fig: filtered_image}. While it appears to be very bright in the filtered image (where the lower limit of the image scaling is set at 3 times the root-mean-square in the blank sky region), it is barely detected in the original F160W image and is completely not detected in the F475W image (right panel of Figure~\ref{fig: filtered_image}). This candidate lies in the high detection probability region, as shown earlier in Figure~\ref{fig: rate_map}, which makes it a possible lensed star candidate, and could be a red supergiant as it is only picked up at F160W but not in F475W (due to the shape of the black body and Balmer break). M25 lens model predicts a lensing magnification of $\sim 100 \pm 2$ at the position of this candidate\footnote{Uncertainty from the standard deviation of 100 realizations following the sampler's posterior in \citet{Melo-Carneiro_2025}}, at the regime that favours lensed star detection \citep{Li_2025_IMF}.
That said, using the same aperture as in Appendix~\ref{sec: appendix_limit}, we carry out photometry on the candidate at the F160W image, and only found a $\sim 2.9\sigma$ significance; the same exercise returns no significant detection at the same position in F475W. The fact that we do not see a transient at F160W is in $1\sigma$ agreement with the predicted detection rate at the same detection limit of $\sim 27.3 m_{AB}$.


Of course, a further caveat is that there are intrinsic variables such as supernovae \citep[SNe, e.g.,][]{Kelly_2015} and luminous blue variables \citep[e.g.,][]{Diego_2024_3M, Li_2024_Flashlights} that can also appear as transients. With two epochs of observation, it would be very challenging to classify between lensed stars and these variables, except for the fact that SNe might appear anomalously bright \citep[as they can be as bright as $M \approx -19$,][]{Richardson_2014} which is not achievable with microlensing of background luminous stars as limited by their radii (see earlier discussion in Section~\ref{sec: method}). Further distinction would then heavily rely on the survey strategy: for example, timely followup observations would allow one to take spectra of lensed star candidates\footnote{subtracting the two spectra taken during when the lensed stars are present, and another one after the lensed stars faded away} and classify the transients based on spectroscopic features; deploying narrow band filters that align with characteristic emission lines can distinguish some classes of intrinsic variables (e.g., SNIa lacks hydrogen lines); of course the most powerful constrain would be to look at the light curve with multi-epoch observations, as microlensing transients and intrinsic variables would have different characteristic light curves \citep[e.g., ][]{Riess_1999, Kelly_2018_Icarus, 2024SSRv..220...57W}. Further discussion is beyond the scope of our work, as we only aim to deliver a prediction for lensed star detection rate in the Cosmic Horseshoe.

All in all, as one's concern could be the statistics of lensed stars being polluted by intrinsic variables, one can also estimate the expected detection of these stars to see if a significant effect is expected, and whether measures in survey strategy should be taken to minimize such a concern. For example, our inferred SFH as shown earlier in Section~\ref{sec: method} would infer a SNIa rate of $\sim 0.1\, \textrm{yr}^{-1}$ \citep{Scannapieco_2005, Sullivan_2006}, and a core-collapse SN rate of $\sim 0.8\, \textrm{yr}^{-1}$ \citep{madau_2014, Chen_2022}. These values are orders of magnitude lower than our predicted lensed star detection rate in the Horseshoe, and thus the contamination should not affect any of the scientific inference from lensed stars statistics mentioned earlier in Section~\ref{sec: applicatoin}.


\subsection{Uncertainties in predicting the lensed star detection rate}
\label{sec: uncertainty}

Here in this subsection, we discuss a couple of caveats in our calculation and address whether they have significantly affected our inference of the transient detection rate in the Cosmic Horseshoe. We shall also discuss future ways of improving the inference whenever possible. 

\subsubsection{Lensing Magnification}

We did not consider the existence of substructures such as subhaloes predicted by hydrodynamical simulations under the cold dark matter paradigm \citep{Moore_1999}. The main reason is that adding subhaloes/density modulation on top of the macro-lens model only perturbs the spatial distribution of transients \citep{Williams_2024, Broadhurst_2025}, whereas the total detection rate over the whole lensed arc remains unchanged \citep[fractional change $\lesssim 1\%$,][]{Palencia_2025_microlensing}. Since we only aim to deliver a prediction of the transient detection rate of the Cosmic Horseshoe under {\it JWST} observations, the fact that the inclusion of subhaloes does not affect the detection rate justifies us in not considering substructures in our calculation.

Another uncertainty comes from the lens model adopted. M25 is the most recent lens model of the Cosmic Horseshoe system, which adopted the most constraints (\textit{HST} imaging of both the Horseshoe and the radial arc, as well as resolved stellar kinematics). This model successfully reproduces almost the entire Horseshoe HST image with eight parameters\footnote{We refer readers to Table 3 in \citet{Melo-Carneiro_2025} for further details.}, accounting for the separate contributions of stellar and dark matter mass. Previous models were either unable to fully reproduce the majority of the Horseshoe {\it HST} image \citep{Bellagamba_2017, Cheng_2019, Schuldt_2019}, or did not adopt the {\it HST} image as a constraint \citep{Belokurov_2007, Dye_2008}. The M25 model should, therefore, be the most reliable model available in terms of both the position of the critical curve (given its success in image reconstruction) and the magnification distribution, which will directly affect the transient detection rate \citep{Li_2025_IMF}. Furthermore, as the model adopts a fully self-consistent approach combining the lens model and the stellar kinematics, degeneracies such as the mass-sheet degeneracy and the mass-anisotropy degeneracy \citep{Gerhard_1993} are removed or at least significantly alleviated. Thus, it provides a more robust assessment of both magnifications and critical curves.



While the M25 lens model has been successful in reproducing most of the features of the Cosmic Horseshoe, there are a couple of small features that are not well reproduced. These sources could be lensed stars, as we have discussed in the last section. But given the expected transient detection rate at the current depth, it is nearly impossible that all of them are lensed stars. The failure in reproducing some of the features, however, should not affect the inference of the detection rate a lot, as the corresponding regions only constitute a very small fraction of the pixels in the arc ($\lesssim 0.1\%$). Even though they are $\sim 10$ times brighter than other pixels, they would only account for $\sim 1\%$ of the detection rate (Equation~\ref{eqn: scale_flux}) and are rather insignificant towards the total rate estimated. This would not affect our prediction of whether transient events could be detected in the Cosmic Horseshoe. The arrival of {\it JWST} data associated with transient observations will simultaneously reveal detailed structures (e.g., star-forming regions) in the Horseshoe that are visible in {\it JWST} but not in {\it HST} imaging. Upon careful removal of detected lensed stars (e.g., modelling and subtracting the lensed stars extracted from differencing the two epochs of images), these detailed structures could also improve the lens model and reduce the remaining uncertainties therein.

Last but not least, GGSL systems are known to suffer from the mass sheet degeneracy -- despite the fact that the position of the critical curve is often well constrained (thus mitigating the concern of defining the spatial distribution of the transient with respect to the critical curve), the underlying lensing potential and therefore magnification distribution in lensed arcs is degenerate with the source position. A different magnification distribution would affect the transient detection rate since they are coupled as a result of magnification bias \citep{Broadhurst_2025, Li_2025_IMF}. For the case of the M25 lens model, the two background sources at different redshifts \citep[the Horseshoe, and the radial arc, see also][]{Schuldt_2019} combining with the fact that it is simultaneously constrained by the stellar kinematics, the mass sheet degeneracy \citep[at least the ``internal'' mass sheet, i.e., mass of the lens, but not those contributed by the external masses such as line-of-sight structures, see][]{Suyu_2010} should have been significantly reduced. 


\subsubsection{Stellar microlensing}

We adopted the stellar component in the M25 lens model to conduct our analysis, which is obtained by isolating the lensing galaxy and fitting the stellar mass freely (by varying the mass-to-light ratio) during the model optimization. Since we have also subtracted the lensing galaxy by fitting Sersic profiles at the different observed bands, we construct the SED of the lensing galaxy and carry out an SED fitting with {\tt Bagpipes} to look for the total stellar mass, independent of the lens model. Assuming a Kroupa IMF and an exponential decay SFH, the best fit total stellar mass of the lensing galaxy is $\sim 8.2\times 10^{11} M_{\odot}$ with a metallicity of $0.2\,Z_{\odot}$. If we assume a constant mass-to-light ratio throughout the lensing galaxy, then the stellar mass distribution of our solution is steeper than that in \citet{Melo-Carneiro_2025}, yielding $\sim 10-30\%$ fewer stellar microlenses (depending on the filter, where F160W yields the largest discrepancy of $\sim30\%$) towards the Cosmic Horseshoe. We recompute the detection rate with this solution, and find an increase in the detection rate by a factor of $\sim 10-30\%$ across all filters and detection limit, reaching $\sim 80$ in F150W for a $5\sigma$ detection limit of $29\,m_{AB}$. This is probably the result of ``more-is-less'' \citep{Diego_2019_extrememagnification, Palencia_2025_microlensing} where abundant stellar microlenses would have saturated the magnification, thus leading to a lower chance of background sources having a higher magnification and eventually leading to a lower detection rate.
While the different adopted value of abundance of stellar microlenses would not affect our conclusion here that lensed star detection is very likely in the Cosmic Horseshoe, future far-infrared {\it JWST} observations are expected to better capture the morphology as well as the abundance of stellar microlenses of the lensing galaxy, and allow us to better probe the abundance of stellar microlenses. This would reduce the uncertainty in detection rate inference and the astrophysical application therein (e.g., probing the IMF).

\subsubsection{Star formation rate}

We mentioned early in Section~\ref{sec: sLF} that we have adopted the emission line EW measured at the brightest part of the Cosmic Horseshoe, instead of the whole arc. This might have overestimated the recent star formation rate; thus, our previous estimation of the transient detection rate could be higher than the true value. To investigate the effect of this uncertainty, we redo the analysis and see how the predicted transient varies with the EW of the emission line narrowed, which should correspond to weaker star formation.

We test with a case where the EW is half the observed value. We found that the inferred SFH remains vastly similar to the original one, predicting a detection rate of $\sim 67 \pm 10$ per pointing in F150W under a detection limit of $29\,m_{AB}$, $1\sigma$ within the value shown earlier in Section~\ref{sec: result}. Looking at the SFH reveals that the decrease in EW suppresses the star formation rate over the last $\sim 10\,$Myr -- the time scale where the strength of these emission lines is most sensitive towards. To get a good fit with the SED, the star formation rate is increased at the last $10-50\,$Myr. Overall, the number of DTM stars remains vastly unchanged and hence explains the similar detection rate.

The only currently available integral-field unit observation on the Cosmic Horseshoe is VLT-MUSE \citep{James_2018, Melo-Carneiro_2025}, which does not cover any of the prominent emission lines associated with star formation (e.g., those discussed earlier in Section~\ref{sec: sLF}) given the redshift of the Cosmic Horseshoe. Future near-infrared integral-field units observation with spatially resolved spectroscopy would cover emission lines, and could mitigate this uncertainty and allow one to better constrain the SFH, and thus predict the transient detection rate.

Another uncertainty comes from the SFH model adopted. Although we have adopted the step-function SFH, which has the highest degree of freedom for inferring SFH, the prior information of the choice of age bins might affect the inferred sLFs. We hence compute also the transient detection rate based on the other two step-function SFH models shown in Appendix~\ref{sec: appendix_SFH}, which have more age bins, thus higher complexity. We found that the transient detection rate remains similar, and well within $1\sigma$ of that shown earlier in Section~\ref{sec: result}.

On top of that, we compute the detection rate while using two other commonly adopted parametric SFH models -- an exponential decay SFH where the star formation peaks at some look-back time and decays exponentially, and a constant SFH where the SFR is constant between some period of look-back time. The former SFH model converges to a similar shape compared with the step-function SFH inferred earlier in Section~\ref{sec: sLF}, where the SFR peaks at $\sim 50\,$Myr ago and has a SFR of $\sim 5\, M_{\odot}\textrm{yr}^{-1}$ at the last couple of Myrs. The constant SFH model, on the other hand, failed to converge. We found that the transient detection rate predicted for the exponential decay SFH is roughly the same (within $1\sigma$) as the one shown earlier in Section~\ref{sec: result}, with a detection rate of $\sim 70$ at F150W with a detection depth of $29\,m_{AB}$. This aligns with the earlier findings in \citet{Li_2025_IMF} that the predicted transient detection rate appears not to be affected by the choice of SFH model. 

We mentioned earlier in Section~\ref{sec: method} that previous literature has estimated a different SFR in the Horseshoe with either UV continuum or H$\alpha$ luminosity, assuming a constant SFR over the last $\sim100\, $Myr. On the other hand, our inference adopted all the available constraints and also has the largest flexibility in the SFH, instead of a constant SFR. To see how the difference in inferred SFR impacts our result, we can naively scale the inferred SFR with the lensed star detection rate, such that even with the lowest SFR estimate \citep[$56\,M_{\odot}\textrm{yr}^{-1}$,][]{Quider_2009}, we are still expecting $\sim 20$ lensed star per visit at F150W under a detection limit of $\sim 29\,m_{AB}$. The conclusion that detecting lensed stars is possible in the Horseshoe holds, but the constraining power of axion mass (if dark matter is composed of ultra-light axions) and stellar IMF would, of course, weaken. Notice that such a naive scaling should not be robust, as the number of transient lensed stars detected per visit per stellar mass drops quickly with increasing look-back time \citep{Li_2025_BRratio} that the relative SFR in the SFH is deterministic towards the analysis. Since our inference has the highest model complexity with the most constraints, we do not repeat our analysis on all the possible SFRs inferred by different works.

\section{Conclusion}

While lensed star transients have been regularly detected in deep observations of galaxy clusters with many interesting astrophysical applications therein, including but not limited to distinguishing the nature of dark matter and probing the stellar initial mass function, there are no detections in galaxy-galaxy strong lensing systems (GGSL) so far. With one of the clear advantages of GGSL over cluster lensing being that there is less room for uncertainty for the lens model, we predict in this work the transient detection rate in a GGSL system and explore the possible astrophysical applications upon transient detection.

We made use of archival {\it HST} observation to predict the {\it JWST} transient detection rate in one of the strongest known GGSL systems -- the Cosmic Horseshoe $(z = 2.381)$. Given the extremely-high recent star formation rate over the last $\sim 50\,$Myr ($\sim 140\,M_{\odot}\textrm{yr}^{-1}$, inferred combining spectral energy distribution and emission line equivalent width) of the Cosmic Horseshoe as inferred in Section~\ref{sec: method}, we predict a transient detection rate of $\sim 60-80$ per pointing in {\it JWST} F150W at a depth of $\sim29\,m_{AB}$ (depending on the choice of SFH model, and the adopted abundance of stellar microlenses), which is the filter that is most sensitive towards the red supergiants associated with the star formation in the past $\sim 50\,$Myr. With this forecast, we highlight two immediate applications:

\begin{itemize}
    \item Previous work has demonstrated with cluster observations that the spatial distribution of transients is sensitive towards the underlying assumed substructures owing to different dark matter models \citep{Williams_2024, Broadhurst_2025}. The high expected transient detection rate of the Cosmic Horseshoe, combined with the fact that there is little room for uncertainty in the position of the critical curve compared with galaxy clusters, makes it a very valuable target to look at the spatial distribution of transients with respect to the critical curve and therefore probe the nature of dark matter. We demonstrated via a simple first-order calculation that a couple of observations of the Cosmic Horseshoe would allow one to distinguish between different Axion masses and place constraints on them.
    
    \item With the distance modulus of $\sim 46.5$ at redshift $2.4$ usually seen as a challenge that limits the transient detection rate, the high star formation rate ($\sim 140\,M_{\odot}\textrm{yr}^{-1}$ as inferred via SED fitting, additionally constrained by equivalent width of four emission lines) of the Cosmic Horseshoe overcomes this obstacle, and it could conversely be seen as a filter to screen out lower mass stars as transients. This enables a unique window looking at the highest mass stellar population at the cosmic noon directly. This is beneficial for studying the most recent star formation, where the light is dominated by the most massive stars. The detection rate would also be extra sensitive towards the slope of the initial mass function, and allow us to place constraints on the initial mass function during the cosmic noon. We predict that with two {\it JWST} observations with $5\sigma$ detection limit of $29\, m_{AB}$, the Cosmic Horseshoe would allow us to tell whether the initial mass function has significantly deviated from the local measurement.
\end{itemize}

The Cosmic Horseshoe is not the only GGSL system with a high recent star formation rate; systems at different redshifts, such as FOR J0332-3557 \citep[star formation rate of $\sim 20 M_{\odot}\textrm{yr}^{-1}$ at $z = 3.8$, ][]{Cabanac_2008, Citro_2024}, could also be perfect candidates to look for transients at different redshift, albeit the fact that they might not reach a high lensed star detection rate as we predict for the Horseshoe. With the upcoming era of sky survey led by Euclid and LSST, one can anticipate the many detections of more relevant GGSL systems \citep[e.g., ][]{Walsmley_2025, Lines_2025}. Deep imaging follow-up of these GGSL systems with {\it JWST} would create a very competent database of transient detection on top of cluster samples, substantially expanding our sample size of lensed star transients at a wide range of source redshifts.

\section*{Acknowledgements}

We sincerely thank the anonymous referee for the many insightful comments that have improved the clarity of this work.
We would like to thank Arsen Levistkiy for the many useful discussions regarding the SED fitting for the Cosmic Horseshoe. 

S.K.L., J.L., and A.C. acknowledge support from the Research Grants Council (RGC) of Hong Kong through the General Research Fund (GRF) 17312122.
L.W., T.E.C., and W.E. acknowledge support from the European Research Council (ERC) under the European Union’s Horizon 2020 research and innovation programme (LensEra: grant agreement No. 945536).
J.M.D. acknowledges support from project PID2022-138896NB-C51 (MCIU/AEI/MINECO/FEDER, UE) Ministerio de Ciencia, Investigaci\'on y Universidades.
C.R.M.C. acknowledges the support of the Conselho Nacional de Desenvolvimento Científico e Tecnológico (CNPq), through grant 140899/2021-9.

We acknowledge the usage of the following programs: Astropy \citep{astropy:2013, astropy:2018, astropy:2022}, Numpy \citep{numpy}, Matplotlib \citep{matplotlib} and Scipy \citep{scipy}.

\section*{Data Availability}


This research is based on observations made with the NASA/ESA {\it Hubble Space Telescope} obtained from the Space Telescope Science Institute, which is operated by the Association of Universities for Research in Astronomy, Inc., under NASA contract NAS 5–26555. These observations are associated with programs 11602 and 12266. 
The data are available at MAST with DOI: 10.17909/0m90-tq84

The authors will share the results generated in this work upon reasonable requests.



\bibliographystyle{mnras}
\bibliography{example} 




\appendix

\section{Choice of step function SFH bins}
\label{sec: appendix_SFH}

\begin{figure}
    \centering
    \includegraphics[width=1\linewidth]{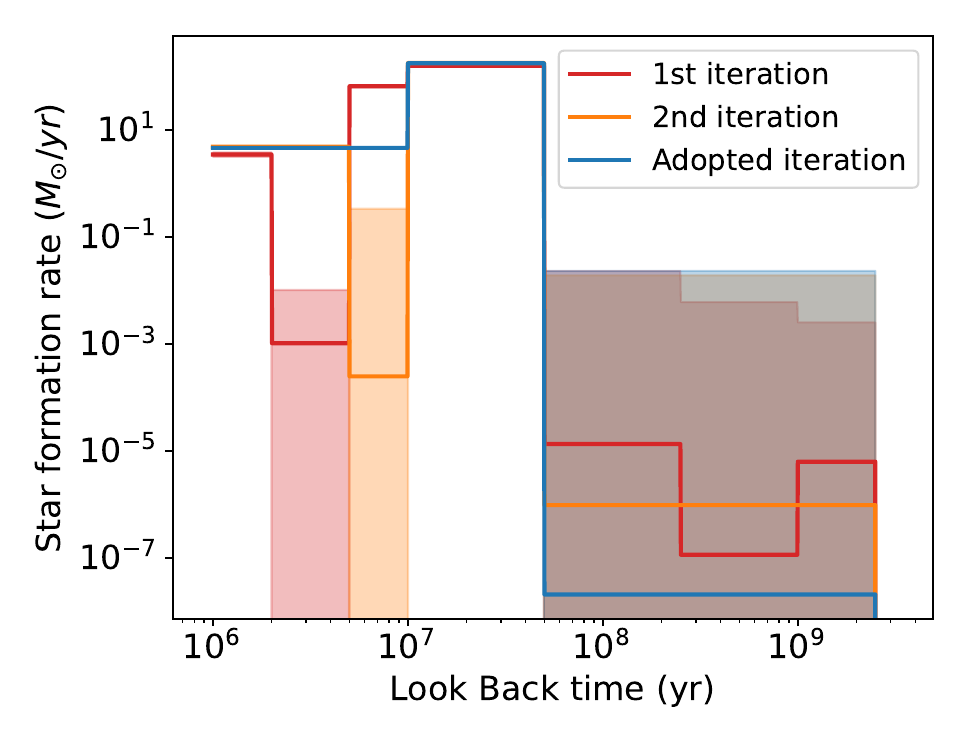}
    \caption{Best-fitting SFH of the Cosmic Horseshoe with different choices of age bins, where the shaded region represents the $\pm1\sigma$ uncertainty. The red SFH is the most complicated step-function SFH we have tried in this work, where we reduce the binning upon combining bins with low signal-to-noise with their neighbour bin(s), arriving at the orange SFH (2nd iteration). Since the second bin (5-10 Myr) of the 2nd iteration has a poor signal-to-noise ratio, we again combine it with the first bin (1-5 Myr) into a single bin of 1-10 Myr and thus arrive at the blue SFH (adopted iteration).}
    \label{fig: sfh_test}
\end{figure}

We begin the SED fitting process by choosing the age bins of $1-2\,$Myr, $2-5\,,$Myr, $5-10\,$Myr, $10-50\,$Myr, $50-250\,$Myr, $250\,$Myr - $1\,$Gyr, and $1-2.5\,$Gyr, similarly to what was done in \citet{Li_2025_IMF, Williams_2025}. The corresponding SFH and the associated $1\sigma$ uncertainty are shown in red in Figure~\ref{fig: sfh_test}.
Since an arbitrarily large number of bins would lead to over-fitting, we then reduce the number of bins by combining bins that have the same star formation rate as their neighbour bins to within $1\sigma$; or when the individual bins have huge uncertainty (signal-to-noise $<<1$, except the last bin since it is expected to be insignificant towards the contribution to DTM stars thus the detection rate), if the quality of fit to the SED (as determined by the reduced $\chi^{2}$) is not decreased.  
Iterating this process, we arrive at the minimum age bins required to get a good fit of the SED being three bins of $1-10\,$Myr, $10-50\,$Myr, and $50\,$Myr$-2.5\,$Gyr, respectively. The corresponding SFH is shown in blue in Figure~\ref{fig: sfh_test}. All SFH can reproduce the observed SED down to a reduced $\chi^{2}$ of $\sim 3$ with a similar recent star formation rate of $\sim 140\,M_{\odot}\textrm{yr}^{-1}$ averaged over the last $50\,$Myr, with the last (simpliest, thus adopted model) having a slightly lower reduced $\chi^{2}$. 

\section{Observed detection limit}
\label{sec: appendix_limit}

We compute the detection limit of the {\it HST} observations used in this paper. This is carried out by injecting fake point sources with known apparent magnitude in empty patches of the sky in each filter, and recovering them with aperture photometry. We adopt a radius of 5 pixels, corresponding to roughly $\sim 2$ times the point-spread-function's full-width half-maximum that captures $\sim 80\%$ of fluxes of point sources, and an annulus of 7 pixels to estimate the local background. The detection limit is thus the dimmest star we can detect with $5\sigma$ confidence, as listed out in Table~\ref{tab: detection_limit}.

\begin{table}
    \centering
    \begin{tabular}{c|c|c}
       Filter  & $5\sigma$ detection limit & Exposure time (s) \\
       \hline
       F275W  & 31.0 & 28100\\
       F475W  & 29.1 & 5500\\
       F606W  & 27.6 & 2400\\
       F814W  & 27.2 & 5600\\
       F110W  & 27.8 & 2400\\
       F160W  & 27.3 & 3800\\
    \end{tabular}
    \caption{Detection limits in HST Cosmic Horseshoe images in $m_{AB}$}
    \label{tab: detection_limit}
\end{table}


\bsp	
\label{lastpage}
\end{document}